\documentclass[12pt]{article}
\usepackage{amsmath} 
\usepackage{xspace} 
\usepackage{booktabs} 
\usepackage{hyperref} 
\hypersetup{colorlinks=true, linkcolor=blue,citecolor=blue}
\usepackage{multirow}
\usepackage{placeins} 
\usepackage{graphicx} 
\usepackage[utf8]{inputenc} 
\usepackage{tikz} 
\usetikzlibrary{calc}
\usepackage[a4paper,margin=3cm]{geometry} 
\usepackage[sorting=none,backend=bibtex]{biblatex}
\usepackage[affil-it]{authblk}
\usepackage[margin=0mm,justification=justified,font=sf,labelfont=bf,singlelinecheck=off,skip=0pt]{caption}
\usepackage{upgreek}
\usepackage{lineno}

\newlength{\plotwidth}
\newcommand{\sLM}{\ensuremath{\sigma_\text{LM}}\xspace}
\newcommand{\sHM}{\ensuremath{\sigma_\text{HM}}\xspace}
\newcommand{\stot}{\ensuremath{\sigma_\text{tot}}\xspace}

\newcommand{\sel}{\ensuremath{\sigma_\text{el}}\xspace}
\newcommand{\sinel}{\ensuremath{\sigma_\text{inel}}\xspace}

\newcommand{\Nel}{\ensuremath{N_\text{el}}\xspace}
\newcommand{\Ninel}{\ensuremath{N_\text{inel}}\xspace}
\newcommand{\Nvis}{\ensuremath{N_\text{vis}}\xspace}

\newcommand{\M}{^M}
\newcommand{\mi}{\text{mi}}
\newcommand{\md}{\text{md}}

\title{
Extraction of low-mass diffractive cross section from the discrepancy between ATLAS and TOTEM total cross sections
}
\author[1]{Per Grafström}
\author[2]{Rafał Staszewski}
\affil[1]{\small Universit\'a di Bologna, Dipartimento di Fisica, 40126 Bologna, Italy}
\affil[2]{\small The Henryk Niewodniczański
Institute of Nuclear Physics
Polish Academy of Sciences,
ul. Radzikowskiego 152,
31-342 Kraków, Poland}

\date{\today}

\begin{document}

\maketitle

\begin{abstract}
At the LHC, two experiments -- ATLAS and TOTEM -- measure the total proton--proton cross section.
Unfortunately, a significant discrepancy, persisting at different collision energies, is observed between the values reported by the two groups.
This paper considers the hypothesis that this tension is predominantly driven by the assumption about the low-mass diffraction used in one of the measurement methods.
It is shown that in such a case it is possible to extract the low-mass diffraction cross section from measurements of ATLAS and TOTEM.
The results are compared with other data-driven estimates, showing better agreement than the original assumption.
\end{abstract}

\section{Introduction}

Diffractive processes are a class of hadron interactions in which quantum-wavelike properties of the projectiles manifest.
These interactions are driven by the strong force and, in proton--proton collisions, constitute a significant fraction of the total cross section.
There are several types of diffractive processes: elastic scattering, diffractive dissociation, central diffractive production.
They are an effect of the Pomeron exchange in the $t$-channel or, in the language of QCD, the exchange of a colour-singlet state of gluons.

\begin{figure}[hbt]
\begin{center}
\newcommand{\xscale}{0.8}
\begin{tikzpicture}[thick,xscale=\xscale]
   \coordinate (A) at (-2, 1.2);
   \coordinate (B) at ( 0, 1);
   \coordinate (C) at ( 2, 1.2);
   \coordinate (a) at (-2, -1.2);
   \coordinate (b) at ( 0, -1);
   \coordinate (c) at ( 2, -1.2);
   \draw[line width = 2pt] (A) node[left] {$p$} -- (B) -- (C) node[right] {$p$};
   \draw[line width = 2pt] (a) node[left] {$p$} -- (b);
   \draw[thick, double, double distance = 2pt] (B) -- (b);
   
   
   \coordinate (d) at (0, 0.18);
   \draw (c) node[right] {$X$};
   \draw (b) -- ($(c) + 0.5*(d)$);
   \draw (b) -- ($(c) - 0.5*(d)$);
   \draw (b) -- ($(c) + 1.5*(d)$);
   \draw (b) -- ($(c) - 1.5*(d)$);
\end{tikzpicture}
\hfill
\begin{tikzpicture}[thick,xscale=\xscale]
   \coordinate (A) at (-2, -1.2);
   \coordinate (B) at ( 0, -1);
   \coordinate (C) at ( 2, -1.2);
   \coordinate (a) at (-2, 1.2);
   \coordinate (b) at ( 0, 1);
   \coordinate (c) at ( 2, 1.2);
   \draw[line width = 2pt] (A) node[left] {$p$} -- (B) -- (C) node[right] {$p$};
   \draw[line width = 2pt] (a) node[left] {$p$} -- (b);
   \draw[thick, double, double distance = 2pt] (B) -- (b);
   
   \coordinate (d) at (0, 0.18);
   \draw (c) node[right] {$X$};
   \draw (b) -- ($(c) + 0.5*(d)$);
   \draw (b) -- ($(c) - 0.5*(d)$);
   \draw (b) -- ($(c) + 1.5*(d)$);
   \draw (b) -- ($(c) - 1.5*(d)$);
\end{tikzpicture}
\hfill
\begin{tikzpicture}[thick,xscale=\xscale]
   \coordinate (A) at (-2, 1.2);
   \coordinate (B) at ( 0, 1);
   \coordinate (C) at ( 2, 1.2);
   \coordinate (a) at (-2, -1.2);
   \coordinate (b) at ( 0, -1);
   \coordinate (c) at ( 2, -1.2);
   \draw[line width = 2pt] (A) node[left] {$p$} -- (B);
   \draw[line width = 2pt] (a) node[left] {$p$} -- (b);
   \draw[thick, double, double distance = 2pt] (B) -- (b);
   
   \coordinate (D) at (0, 0.12);
   \draw (C) node[right] {$Y$};
   \draw (B) -- ($(C) + 0.5*(D)$);
   \draw (B) -- ($(C) - 0.5*(D)$);
   \draw (B) -- ($(C) + 1.5*(D)$);
   \draw (B) -- ($(C) - 1.5*(D)$);
   
   \coordinate (d) at (0, 0.18);
   \draw (c) node[right] {$X$};
   \draw (b) -- ($(c) + 0.5*(d)$);
   \draw (b) -- ($(c) - 0.5*(d)$);
   \draw (b) -- ($(c) + 1.5*(d)$);
   \draw (b) -- ($(c) - 1.5*(d)$);
\end{tikzpicture}
\end{center}
\caption{\sf Pomeron exchange diagrams corresponding to single and double dissociation.}
\label{fig:diagrams}
\end{figure}
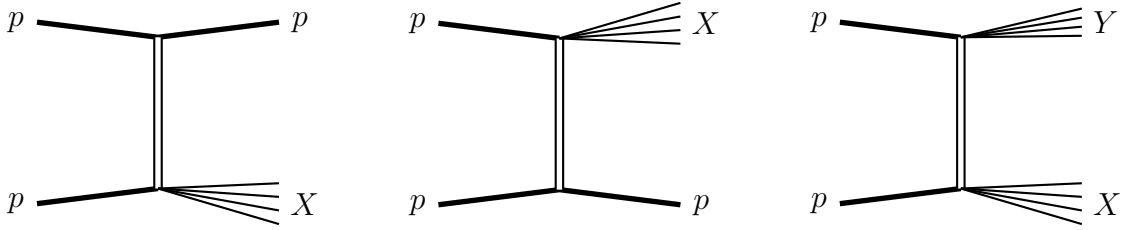

The diagrams of diffractive dissociation interactions in $pp$ collisions are presented in Fig. \ref{fig:diagrams}.
The first two show single diffractive dissociation, commonly called single diffraction (SD), where one of the interacting protons dissociates into a multi-particle state $X$.
When both protons dissociate, the process is called double diffractive dissociation or double diffraction (DD).

Because of the intrinsically non-perturbative nature of diffractive dissociation, the possibilities of their theoretical descriptions are limited.
For low masses, \textit{i.e} $M_X$ and $M_Y$ below 3--4 GeV, the Good--Walker \cite{Good:1960ba} formalism is often used.
Here, the proton state is described as a superposition of several eigenstates of the elastic scattering matrix.
As a consequence, elastic interaction leads to processes with dissociation.
The Good--Walker approach is practical at low masses because here it is reasonable to consider only a few effective eigenstates, which results in a limited number of free parameters of the model.

The high-mass regime is instead typically modelled within the Regge theory (see \textit{e.g.} \cite{Kaidalov:1973tc,Collins:1977jy}).
The description uses the triple-pomeron coupling, which leads to a smooth $1/M_X^2$ distribution.
Because of this, it is not applicable for low masses, which are dominated by resonances.

Diffractive dissociation is not only hard to describe theoretically but also difficult to measure at high-energy colliders.
This difficulty originates from the fact that the scattered protons and many of the particles of the dissociated state are emitted in the very forward direction and are lost in the beam pipe.
While scattered protons can be registered using the Roman-pot technique,
the low-mass dissociated systems can easily escape completely undetected.

Diffractive dissociation has of course an interest in itself but is also related to the measurements of the total cross section, \stot, at LHC.
This is becaue estimates of the cross section for low-mass diffraction, \sLM, are an essential ingredient in the \stot measurements performed by the TOTEM Collaboration.
Actually, the uncertainties assigned to \sLM constitute the largest contribution to \stot uncertainty for TOTEM's main experimental method.

Results on \stot are also available from the ATLAS Collaboration.
ATLAS used a different method, which did not rely on any assumptions about the diffractive dissociation.
However, contrary to TOTEM's method, it required a dedicated luminosity measurement.
Both groups measured \stot at centre-of-mass energy of 7, 8 and 13~TeV.
Since the first result at 7 TeV, a tension between the results of the two groups has been observed.

In this paper, we offer an explanation of the observed tension. 
The explanation is based on an assumption that all experimental procedures of both ATLAS and TOTEM are correct.
This implies that the tension must originate from the theoretical input.
The obvious suspect is the unmeasured low-mass diffraction cross section.

The plan of the paper is as follows.
First, in Sec. \ref{sec:stot}, we present the state of the art for the measurements of total cross sections at LHC.
The different methods of measurement are introduced and the available results are summarised.
In Sec. \ref{sec:interpretation}, we explain the hypothesis that the source of the tension is the assumption about the low-mass diffraction.
We show that if the hypothesis is correct, it is possible to derive the true value of the low-mass diffraction cross section by combining the measurements of ATLAS and TOTEM.
In Sec. \ref{sec:results} we present the resulting values.
Since the results of such a procedure depend solely on data, \textit{i.e.} data from ATLAS and TOTEM, they can be considered as measurements of low-mass diffractive dissociation.
In Sec.~\ref{sec:others}, we provide a test for the hypothesis by confronting its consequences with other LHC measurements sensitive to low-mass diffraction.

\section{Measurement of the total cross section at LHC and the different methods applied}
\label{sec:stot}

The measurement of the total $pp$ cross section is based upon the optical theorem, which relates its value to the imaginary part of the forward scattering amplitude:
\begin{equation}
\label{eq:optical_theorem}
    \stot  \propto  \text{Im}\, f_\text{\bfseries el}(t=0).
\end{equation}
The standard procedure is to assume that the phase of the amplitude is known, which leads to the formula relating the total cross section to the differential elastic cross section:
\begin{equation}
    \label{eq:stot}
               \stot^2 = \left. \frac{16 \pi}{1+\rho^2}\frac{\text{d} \sigma_\text{el}}{\text{d} t} \right|_{t\to 0},
\end{equation}
where the $\rho$ parameter is defined as :
\begin{equation}
\label{eq:rho}
\rho = \frac%
{\text{Re} f_\text{\bfseries el}(t=0)}
{\text{Im} f_\text{\bfseries el}(t=0)}.
\end{equation}

Several methods have been used to measure the total cross section at LHC.
The most straightforward one, the one used for almost all other measurements at LHC, depends on luminosity measurements.
Knowing the integrated luminosity $L$ corresponding to the collected dataset allows a direct conversion of the number of observed events to the cross section value.
This leads to formula
\begin{equation}
    \label{eq:lumidep}
               \stot^2 = \left. \frac{16 \pi}{1+\rho^2} \frac{1}{L} \frac{\text{d} N_\text{el}}{\text{d} t} \right|_{t\to 0}.
\end{equation}

The luminosity-dependent method was used for all ATLAS measurements \cite{A7, A8, A13} and for one of TOTEM at 7 TeV \cite{T7Ld}.
The advantage of this method follows from the fact that the left-hand side of Eq. (\ref{eq:lumidep}) has \stot squared.
Because of this, the uncertainty on \stot behaves approximately as half of the uncertainty affecting the right-hand side of Eq. (\ref{eq:lumidep}).
For example, if the uncertainty on luminosity is 2\%, it will translate into 1\% uncertainty on \stot.

The disadvantage of this method is that it requires a dedicated luminosity measurement,
which may not be easy for conditions used during elastic-scattering data taking.
These conditions are very different from those during standard LHC operations,
which is a consequence of the special settings of the accelerator magnets, so-called machine optics, used for elastic measurements.

A particular machine optics is often referred to by the value of the betatron function at the interaction point, $\beta^\ast$.
Typical LHC operations are characterised by $\beta^\ast$ of few tens of centimetres, which leads to small beam sizes (order of $10~\upmu\text{m}$), high instantaneous luminosity (order of $10^{34}\text{cm}^{-2}\text{sec}^{-1}$) and large angular divergence of beams (about $30~\upmu\text{rad}$).
In order to measure elastic scattering at very small angles, $\beta^\ast$ is increased to 90~m or even 2500~m,
resulting in an increase of the beam size (order $500~\upmu\text{m}$), reduction of luminosity (order of $10^{27}\text{cm}^{-2}\text{sec}^{-1}$), and reduction of the beam divergence (fractions of a microradian).

An alternative method of \stot measurements, deployed for the first time at ISR \cite{ISRluin}, does not use a luminosity measurement, but is based on counting inelastic events in the same data-taking period as the elastic scattering events are collected.  
The method is possible due to the squared \stot in the left-hand side of Eq.  (\ref{eq:lumidep}).
Substituting one of those \stot with $N_\text{tot}/L = (N_\text{el} + N_\text{inel})/L$ leads to formula
\begin{equation}
    \label{eq:lumiind}
            \stot = \left. \frac{16 \pi}{1+\rho^2} \frac{1}{N_\text{el} + N_\text{inel}}
               \frac{\text{d} N_\text{el}}{\text{d} t} \right|_{t\to 0}.
\end{equation}
The disadvantage of this approach is the need to measure \Ninel.
This requires a dedicated detector capable of measuring inelastic interactions.

In practice, in an accelerator experiment, it is impossible to construct a detector that can measure all inelastic collisions.
The process of a particular difficulty is the low-mass diffractive dissociation
because of final-state particles produced at very large rapidities.
In some events, all those particles go into the accelerator beam pipe and cannot be registered by the existing instrumentation.
Thus, only a part of \Ninel can be detected and a correction for the unmeasured events has to be applied.
This method is the main one used by the TOTEM Collaboration in their measurements \cite{T7, T8, T13}.

The third method is based on the fact that the Coulomb part of the elastic scattering is reliably described by QED.
Then, measuring elastic scattering in the Coulomb region can provide normalisation for the full $t$ spectrum. 
This method was employed in the TOTEM measurement at 13 TeV \cite{T13Ld}.

\begin{table}
\caption{Summary of model-dependent (luminosity-independent) measurements.}
\label{tab:md}
\begin{tabular}{lrr}
\toprule[2pt]
                              & $\stot^\md$ [mb] & $\sinel^\md$ [mb]\\
\midrule[1.3pt]
 TOTEM 7 TeV \cite{T7} & $  98.0  \pm   2.5$ &$  72.9  \pm  1.5 $\\
 TOTEM 8 TeV \cite{T8}                 & $ 101.7  \pm   2.9$ &$  74.7  \pm  1.7 $\\
 TOTEM 13 TeV \cite{T13}            & $ 110.6  \pm   3.4$ &$  79.5  \pm  1.8 $\\
\bottomrule[2pt]
\end{tabular}

\vspace{3ex}

\caption{Summary of model-independent measurements.}
\label{tab:mi}
\begin{tabular}{clcccc}
\toprule[2pt]
                                   &  &   $\stot^\mi$ [mb] &   $\sinel^\mi$ [mb]& $R^\mi$ & $B^\mi$ [GeV$^{-2}$]\\
\midrule[1.3pt]
\multirow{3}{*}{\rotatebox{90}{7 TeV}} &
 ATLAS \cite{A7}       & $   95.35   \pm  1.36  $  &  $   71.34   \pm   0.82 $ &$   0.253 \pm  0.005   $ & $   19.73 \pm  0.3     $ \\
 & TOTEM$^a$  \cite{T7Ld} & $   98.3    \pm  2.8   $  &  $   73.5    \pm   1.9 $ &$   0.253 \pm  0.005   $ & $   19.73 \pm  0.3     $ \\
 & Average           & $   95.9    \pm  1.2   $  &  $   71.68 \pm   0.75$  &$   0.253 \pm  0.004$ & $   19.73 \pm  0.2$ \\
\midrule[0.5pt]
\multicolumn{2}{l}{8 TeV ATLAS \cite{A8}}                & $   96.07   \pm   0.92 $  &  $   71.73   \pm   0.71    $ &$   0.252 \pm  0.004   $ & $   19.74 \pm  0.24    $ \\
\midrule[0.5pt]
\multirow{3}{*}{\rotatebox{90}{13 TeV}} &
 ATLAS \cite{A13}                & $  104.7    \pm   1.1  $  &  $   77.41   \pm   1.09    $ &$   0.257 \pm  0.012   $ & $   21.14 \pm  0.13    $ \\
 & TOTEM$^b$ \cite{T13Ld} & $  109.3    \pm   3.5  $  &    --        & --  &   --    \\
 & Average            & $  105.1   \pm   1.0  $  &    --       & --  &   --    \\
\bottomrule[2pt]
\end{tabular}
\\[3pt]
\raggedright
$^a$ luminosity-dependent method \\
$^b$ Coulomb normalisation
\end{table}

In the context of the low-mass diffraction, it is useful to divide the available \stot measurements into two groups.
The first one contains results that depend on model-based assumptions.
These are the results based on the second method discussed above, \textit{i.e.} the main method used by TOTEM.
In the following, measurements from this group will be called \textit{model dependent} or \textit {MC dependent} and denoted by superscript \textit{md}.

The second group contains measurements that do not rely on theoretical assumptions about low-mass diffraction.
They come from the first (luminosity-dependent) and the third (Coulomb normalisation) method.
These results will be called \textit{model independent} or \textit{MC independent} and denoted by superscript \textit{mi}.

A summary of the model-dependent (luminosity-independent) results is presented in Tab. \ref{tab:md}.
Tab. \ref{tab:mi} shows results of the model-independent measurements.
The tables do not show all observables reported in the corresponding papers, but only those that are used in the following calculations.
In Tab. \ref{tab:md} these are \stot and
\sinel{} -- the inelastic cross section.
Tab. \ref{tab:mi} contains also
$R$ -- the ratio of elastic to total cross section, and 
$B$ -- the elastic slope.
The tables do not show TOTEM's measurement in the CNI region at 8~TeV~\cite{TOTEM:2016lxj}.
This result does not provide any additional information about the normalization, since for this measurement the elastic distribution was simply normalized to the distribution obtained from the subset of data used in measurement \cite{T8}.

At 7 and 13 TeV, but not at 8 TeV, model-independent results are available from both ATLAS and TOTEM, and are in good agreement. 
Since they come from different experiments and were obtained using different methods, their uncertainties can be assumed to be independent.
One can then combine these measurements by calculating their weighted average.
These average values are the best estimates of the model-independent results that are available.
They are shown in Tab. \ref{tab:mi} and are used as the default in the following calculations.

\begin{figure}[htpb]
    \centering
    \includegraphics[width=\plotwidth,page=2]{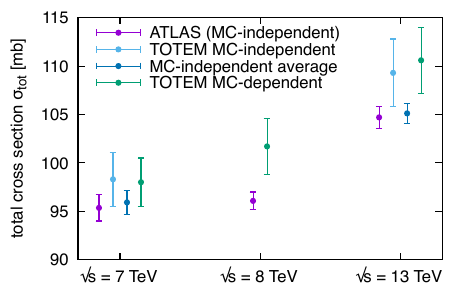}
    \caption{Total $pp$ cross section values measured at LHC.}
    \label{fig:measurements}
\end{figure}

All values of the total cross section are presented in Fig. \ref{fig:measurements}.
Here, one can clearly observe the tension between the ATLAS MC-independent results and the TOTEM model-dependent ones.
This difference can be quantified with the p-value, which for 7, 8 and 13 TeV are 0.35, 0.06 and 0.10, respectively, or with $\chi^2$ values: 0.87, 3.42, 2.73. 

It is important to point out an argument that one may raise, namely that TOTEM's model-dependent \stot values agree well with their model-independent results.
Indeed, the p-values between two TOTEM measurements are very large: 0.94 at 7 TeV and 0.79 at 13 TeV ($\chi^2 = 0.006$ and 0.07, respectively).
However, the uncertainties of TOTEM's model-independent results are also rather large. 
This makes the TOTEM MC-independent results in agreement not only with TOTEM MC-dependent ones, but also with ATLAS.
The p-values corresponding to the latter comparison are 0.34 at 7 TeV and 0.21 at 13 TeV ($\chi^2 = 0.9$ and 1.6, respectively).
Therefore, when one considers the model-independent result from TOTEM as a validation for their model-dependent method, one should have in mind the accuracy of such a validation.
This accuracy is not sufficient to discriminate the TOTEM model-dependent method against the ATLAS model-independent method.

\FloatBarrier

\section{Interpretation of ATLAS--TOTEM tension}
\label{sec:interpretation}

Below, we follow the consequences of the assumptions of the paper:
that all experimental procedures, both in ATLAS and TOTEM have been performed correctly.
Such procedures include various aspects that do not depend on models of the physics being studied, for example: correction for geometric acceptance, correction for trigger and reconstruction efficiency, optics determination, unfolding of resolution effects, \textit{etc.}
Our assumption means, in particular, that all uncertainties related to the above ingredients (including ATLAS luminosity measurements, which are sometimes questioned) are properly estimated.
As anticipated before, the suspected source of the discrepancy becomes the assumption about the fraction of the low-mass diffraction, which is used in the TOTEM's main method.

TOTEM measured the inelastic yields using their T2 detectors \cite{TOTEMJINST,Antchev_2013}.
The acceptance of these detectors reached pseudorapidities of 6.5.
Therefore, events where all charged particles are produced with $|\eta| > 6.5$ cannot be registered.
In particular, diffractive events with a mass of the dissociated system(s) of not more than few GeV have a good chance of not being observed.
This effect is quantified using the $\epsilon$ parameter defined as:
\begin{equation}
N_\text{all inelastic} = (1 + \epsilon) N_{|\eta| < 6.5},
\end{equation}
where $N_{|\eta| < 6.5}$ is the number of events where at least one charged particle with $|\eta| < 6.5$ was produced.
The values of the $\epsilon$ parameter were obtained by TOTEM using the QGSJET-II-03 MC generator \cite{Ostapchenko:2004ss} that was scaled to reproduce the fraction of events with particles observed in the T2 detectors only on one side of the interaction point.
This rescaling, to some extent, constrains high-mass single diffraction and double diffraction with one diffractive mass being low and the other one high.
In the following, the name \textit{TOTEM QGSJET} will be used referring to the predictions including the rescaling.

The correction values used by TOTEM at 7, 8 and 13 TeV are respectively: 0.042, 0.048, 0.071.
Acknowledging that the $\epsilon$ values are known with limited accuracy, TOTEM has set the uncertainty to be 50\% of the value of $\epsilon$, assuming this to be a conservative estimate.
However, TOTEM did not discuss what this estimation was based on.
Since this is an essential ingredient in the measurement and it was not properly justified, we treat the 50\% uncertainty as an arbitrary value.
On the other hand, no obvious alternative was available, since none of the existing models can be trusted without being tuned to high-energy data sensitive to low-mass diffraction.

The present paper attempts to obtain estimations of low-mass diffraction that are based purely on the available data, namely the ATLAS--TOTEM tension.
The basic idea is understood by looking at Eq. (\ref{eq:lumiind}).
The goal is to find what value of \Ninel would have to be used in the model-dependent measurements to obtain the same \stot as in the model-independent ones.

Technically, this can be obtained in several different ways.
Here, we consider four different approaches, A, B, C and D, resulting in four different formulas.
All of them have the same basic structure:
\begin{equation}
\epsilon = \epsilon_0 + \epsilon',
\end{equation}
where $\epsilon$ is the model-independent value obtained in the present paper,
$\epsilon_0$ is the value used by TOTEM,
$\epsilon'$ is the correction to $\epsilon_0$.
The different formulae for $\epsilon'$ are presented in Table \ref{tab:formulas}, 
see Appendix~\ref{sec:deriv} for their derivations.

\begin{table}[h]
\caption{Formulae for $\epsilon'$ in approaches A, B, C and D.}
\label{tab:formulas}
\centering
\vspace{2ex}
\begin{tabular}{c  c p{3ex}  l}
\toprule
\textbf{Approach} & \textbf{Input} &&  \bfseries Formula for $\epsilon'$\\ \midrule
A & $\stot^\md, \stot^\mi, \sinel^\mi$ & &
$\displaystyle \frac{\stot^\md - \stot^\mi}{\sinel^\mi} $
\\[3ex] 
B & $\stot^\md, \stot^\mi, \sinel^\md$ & &
$\displaystyle \frac{\stot^\md \left(\stot^\md - \stot^\mi\right)}{\sinel^\md \stot^\mi}$
\\[3ex] 
C & $\stot^\md, R^\mi, \sinel^\mi$ & &
$ \displaystyle \frac{\stot^\md}{\sinel^\mi} - \frac{1}{1 - R^\mi} $ 
\\[3ex] 
D & $\stot^\md, \stot^\mi, B^\mi$ & &
$\displaystyle \left(\frac{\stot^\md}{\stot^\mi} - 1\right) / \left(1 - \frac{\stot^\mi \left(\rho^{2} + 1\right)}{16 \pi B^\mi }\right)$
\\[3ex] 
\bottomrule
\end{tabular}
\end{table}

The difference between the different approaches comes from the different inputs used for the $\epsilon'$ calculation.
One of those input values is always the model-dependent total cross section value $\stot^\md$, for which the $\epsilon_0$ value was assumed.
At least one of the remaining inputs is coming from the model-independent measurement: either $\stot^\mi$ or $R^\mi$ -- the ratio of elastic to total cross section.
The third input value is: $\sinel^\mi$ (formula A and C), $\sinel^\md$ (formula B) or $B^\mi$ (formula D). 

Comparing different formulae is particularly important in the context of error propagation.
For example, when deriving the uncertainty on $\epsilon'$ using formula A, one should in principle take into account the correlation between the uncertainties of $\stot^\mi$ and $\sinel^\mi$.
This correlation can be sizeable, since both these values originate essentially from the same differential elastic cross-section distribution.
However, the information about this correlation is not publicly available and cannot be used in the present paper.

Other formulae have their own deficiencies.
Estimation of uncertainty based on Formula B is sensitive to correlations between $\stot^\md$ and $\sinel^\md$.
One could expect that the problem of such a correlation is reduced for Formula C, since in the ratio $R^\mi = \sel^\mi / \stot^\mi$ the main source of normalisation uncertainty can, at least partially, cancel.
The drawback of formula C is that the $R^\mi$ value  is not available for all studied samples.
Finally, formula D should also have a smaller, or at least different and independent, problem with correlations.
However, its derivation neglects the non-exponential shape of the differential cross section.

\begin{figure}[htpb]
    \centering
    \includegraphics[width=1.25\plotwidth]{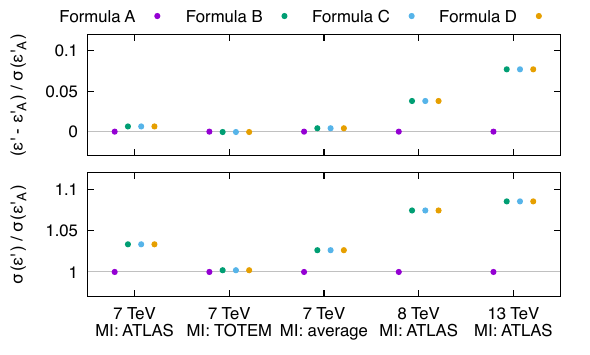}
    \caption{Comparison of the $\epsilon'$ values obtained with different formulae. For each energy and each choice of model-independent set of observables, all points are normalised to the results of formula A.}
    \label{fig:formulas}
\end{figure}

Figure \ref{fig:formulas} presents a comparison of the $\epsilon'$ values obtained for the four different approaches at the collision energies available at the LHC.
At 7 TeV, three model-independent inputs have been used \textit{i.e.} the one of ATLAS, the one of TOTEM and the average of the two.
At 13 TeV, the results using the TOTEM Coulomb normalisation and the model-independent average are not shown, because the available data allows only for Formula B to be applied.
The uncertainty calculation is discussed in Appendix \ref{sec:errorprop}.

It is interesting to observe that despite the differences, all formulae give very similar results, both in value as well as the uncertainty.
While Formula A tends to have a slightly smaller uncertainty, the effect is tiny.
One can then safely conclude that the choice of the approach is irrelevant for the obtained results.
Thus, all considerations in the following will use only Formula B, because it is the only one available for all considered cases.

Typically, the greatest contribution to the $\epsilon'$ uncertainty comes from the error on $\stot^\md$.
However, uncertainties of $\stot^\md$ have a large contribution from the 50\% uncertainty assumed on $\epsilon_0$.
In our derivation,  $\epsilon$ is the true correction that should have been applied, hence, by definition, it cannot depend on the choice of $\epsilon_0$.
Therefore, $\epsilon_0$ has to be treated as a fixed number and should not be attributed any uncertainty.
This applies to the $\epsilon_0$ contribution to the $\stot^\md$ uncertainty.
For this reason, in the above calculation of the $\epsilon'$ uncertainty, the $\epsilon_0$ uncertainty is subtracted from the $\stot^\md$ uncertainty as explained in Appendix~\ref{sec:errorprop}.

\section{Results}
\label{sec:results}

\begin{figure}[htbp]
    \centering
    \includegraphics[width=\plotwidth,page=2]{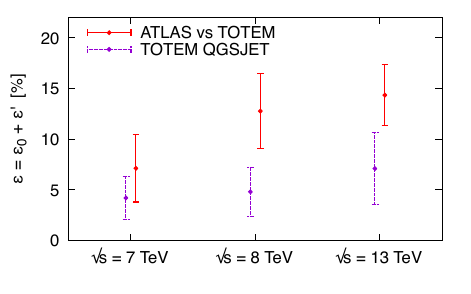}%
    \caption{Comparison of the $\epsilon$ values used by TOTEM with the best estimates from the present paper.
    The errors on the TOTEM QGSJET values are drawn with a dashed line to indicate that they are not intrinsic uncertainty of the model, but an arbitrary assumption.
    }
    \label{fig:results}
\end{figure}

The central results of the analysis are presented in Fig. \ref{fig:results},  where the $\epsilon$ values used by TOTEM are confronted with the values calculated using the best estimate of model-independent quantities available.
At 7 and 13 TeV, these best values are the averages between ATLAS and TOTEM model-independent measurements.
At 8~TeV, only the ATLAS measurement is available.

At 7 TeV, our data-driven $\epsilon$ value is in good agreement with $\epsilon_0$.
At 8~TeV and 13~TeV, our estimate is significantly larger.
Treating the $\epsilon_0$ as a model prediction, \textit{i.e.} as a number without uncertainty, the disagreement with our estimates is $0.9\sigma$, $2.2\sigma$, and $2.4\sigma$ at 7, 8 and 13 TeV, respectively.
The $\epsilon_0$ values are thus incompatible with our results.  
On the other hand, taking into account also the arbitrarily assumed $\epsilon_0$ uncertainty of $\epsilon_0/2$, this tension reduces to $0.5\sigma$, $1.3\sigma$, and $1.1\sigma$.
We can conclude that the 50\% uncertainties on $\epsilon_0$ are at least a little underestimated.

\begin{figure}[htbp]
    \centering
    \includegraphics[width=\plotwidth,page=1]{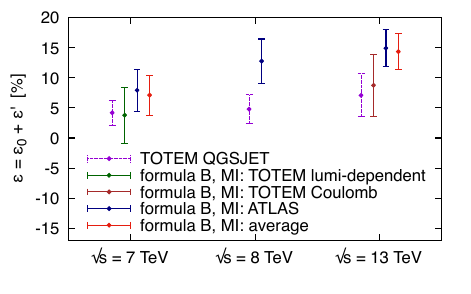}%
    \caption{Comparison of $\epsilon$ values used by TOTEM and obtained with different choices of model-independent results.
    The errors on the TOTEM QGSJET values are drawn with a dashed line to indicate that they are not intrinsic uncertainty of the model, but an arbitrary assumption.
    }
    \label{fig:results_all}
\end{figure}

The full spectrum of the results, including all available choices of model-inde\-pen\-dent results, is shown in Fig. \ref{fig:results_all}.
One can see that the tension between $\epsilon$ and $\epsilon_0$ is mainly driven by the ATLAS measurements.
The results based on the TOTEM model-independent data are in good agreement with the $\epsilon_0$ values.
However, the uncertainties of the TOTEM model-independent inputs are significantly larger than those in the ATLAS case.
This makes the weighted average dominated by the latter input.

Knowing the $\epsilon$ value, one can translate it into the low-mass diffraction cross section.
The conversion is straightforward :
 \begin{equation}
    \label{eq:sd1}
     \sLM = \epsilon \cdot \sinel^\mi.
 \end{equation}
However, for the 13 TeV average model-independent result, $\sinel$ is not available.
Therefore,  the following formula is used instead:
 \begin{equation}
    \label{eq:sd2}
     \sLM = \epsilon \cdot \sinel^\md \cdot \frac{\stot^\mi}{\stot^\md}.
 \end{equation}
While it is not fully equivalent to Eq. (\ref{eq:sd1}), it is a good approximation.
This was confirmed by the results at 7 and 8 TeV, where both formulae lead to very similar values of $\sLM$ and its uncertainty.

Some caution is needed when interpreting the   $\sLM$ values.
Strictly speaking, they correspond to events in which no charged particles with $|\eta| < 6.5$ are produced.
Such events are dominated by low-mass diffraction, but there may be a small contribution of other events.
Non-diffractive processes can lead to low particle multiplicity, and all charged particles may happen to be produced in the forward direction.
In addition, both single and double dissociation processes with low diffractive mass can contribute, but it is expected that the double-diffractive contribution is small \cite{Khoze:2014aca}.

It is worth pointing out that the $|\eta| > 6.5$ condition can be translated to the limit on diffractive mass: $M_X < M_0$.
However, this limit is not sharp, but rather a smooth transition (cf. Fig. 3 of \cite{Antchev_2013}). 
Defining the $M_0$ value by the point of 50\% acceptance leads to values of about 3.4 GeV and 4.6 GeV at 7/8 and 13 TeV, respectively.
Neglecting the smooth transition could be a valid approximation in a situation when the $M_X$ distribution is approximately flat in the transition range, but this is not the case.

\begin{figure}[htbp]
    \centering
    \includegraphics[width=\plotwidth,page=3]{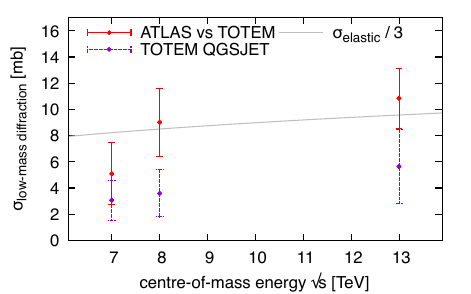}%
    \vspace{5pt}
    \caption{Comparison of the low-mass diffraction cross section used by TOTEM with the results obtained in this paper.
    For the gray line, corresponding to one third of \sel, a parametrization of $\sel(s)$ from Fig. 24 of \cite{A13} was used.
    The errors on the TOTEM QGSJET values are drawn with a dashed line to indicate that they are not intrinsic uncertainty of the model, but an arbitrary assumption.
    }
    \label{fig:results_sigma}
\end{figure}

The estimated cross section values are  presented in Fig. \ref{fig:results_sigma} and Tab. \ref{tab:results_sigma}. 
These are the main numerical results of the present paper.
The conclusions of the comparison with the TOTEM QGSJET are the same as from the $\epsilon$ value.

In the next section, we will compare our result with other measurements at the LHC. 
Here, we point out an intriguing observation when our estimates are compared with the naive expectation from data at ISR and energies below. 
At those centre-of-mass energies, the low-mass diffraction was typically one third of the elastic cross  section (see \cite{Khoze_2018} and references therein \textit{e.g.} \cite{BAKSAY1975484,WEBB1975331,DEKERRET1976477,MANTOVANI1976471}).
A priori, there is no reason that this tendency should continue up to LHC energies.
However, our values of \sLM agree very well with one third of elastic cross section, as can be seen in Fig.~\ref{fig:results_sigma}.

\begin{table}[htbp]
\caption{Data-driven values of cross section for low-mass diffraction.}
\label{tab:results_sigma}
\vspace{2ex}
\centering
\begin{tabular} {c | c c c}
\toprule
centre-of-mass energy, $\sqrt{s}$ & 7 TeV & 8 TeV & 13 TeV \\
\midrule
low-mass diffraction cross section & 
$5.1 \pm 2.4$ mb&
$9.0 \pm 2.6$ mb&
$10.9 \pm 2.3$ mb \\
\bottomrule
\end{tabular}
\end{table}

\FloatBarrier
\section{Comparison with other measurements at LHC}
\label{sec:others}

In this section, we compare our results with other available data.
As mentioned earlier, direct detection of low-mass diffraction is very difficult and no such measurements are available at LHC.
But it is possible to put some constraints using indirect methods that are independent from the analysis presented in this paper.

At 7 TeV, TOTEM measured the elastic differential cross section using the luminosity-dependent method.
One of the outcomes of the measurement is the value of the total inelastic cross section \sinel, which was obtained by subtracting the total elastic cross section from the total cross section.
Such a \sinel value covers all inelastic processes, including the low-mass diffraction.
Using the same data, TOTEM also measured the inelastic cross section with charged particles registered in their T2 detectors.
The difference between these two cross sections constrains low-mass diffraction:
$ \sLM = 2.62 \pm 2.17 $~(mb) ($\sLM < 6.31$~mb at 95\% CL) \cite{TOTEM:2013dtg}.

At 13 TeV, TOTEM did not measure \sinel in a fiducial volume excluding the low-mass diffraction.
On the other hand, similar measurements were performed by ATLAS and CMS \cite{ATLAS:2016ygv,CMS:2018mlc}, but excluding a wider range of masses, \textit{i.e.}  $M_X < 13$~GeV.
There is a very good agreement between the two experiments.
Combining those fiducial cross sections with the total inelastic cross section measured at 13 TeV, one can obtain an estimate of the cross section for a mixture of low-mass diffraction and a part of the high-mass diffraction: 
$\sLM + \sHM = 10.0 \pm 1.4$~mb.
However, in addition there are data from the CMS experiment in the range  $4.1\ \text{GeV} < M_X < 13 \ \text{GeV}$   measuring a cross section of $\sHM = 2.2\pm{0.7}$ mb \cite{CMS:2018mlc}.
By subtracting this contribution we get a cross section for low-mass dissociation of $\sLM = 7.8\pm{1.4}$ at 13 TeV (this argument has also been developed in \cite{lowmassnote} and a similar one in \cite{Khoze_2018}).

\begin{figure}[htpb]
    \centering
    \includegraphics[width=\plotwidth]{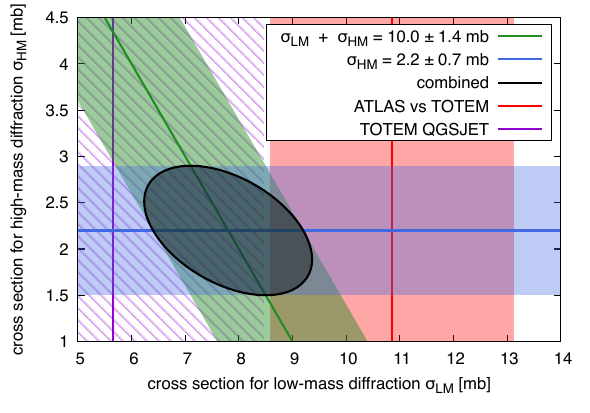}
    \caption{
    Comparison of various constraints of low-mass diffraction at $\sqrt{s}=13$~TeV.
    The bands represent the uncertainty.
    The error on the TOTEM QGSJET value is drawn with a hatched pattern to indicate that it is not an intrinsic uncertainty of the model.
    }
    \label{fig:vis13TeV}
\end{figure}

The argument is illustrated in Fig. \ref{fig:vis13TeV}.
The final constraint is represented by the ellipse.
The figure also shows the result based on the ATLAS--TOTEM discrepancy.
It can be observed that it has a certain overlap with the right side of the ellipse.
One can also see that the TOTEM QGSJET value, if treated as a model prediction -- without uncertainty, is outside the combined constraint.
Taking it with the arbitrary 50\% uncertainty makes it in agreement with the combined result.

A reasoning similar to the one above can also be applied at 7 TeV. 
This way provides an additional constraint at that energy, but without relying on the TOTEM measurements mentioned above \cite{TOTEM:2013dtg}.
Here, the initial situation is similar to 13~TeV: ALICE, ATLAS and CMS have measured the fiducial cross section excluding diffraction masses $M_X$ below 15.7 GeV \cite{ALICE:2012fjm,ATLAS:2011zrx,CMS:2012gek}, and the results of the three experiments are in very good agreement: 62.1 ± 2.4 mb, 60.3 ± 2.1 mb and 60.2 ± 2.6 mb, respectively.
In the following calculations, we take the ATLAS value, which happens to be the middle one and the one with the smallest uncertainty.
This gives $\sLM + \sHM = 11.0\pm2.1$~mb.

At 7 TeV, contrary to the 13 TeV case, there is no direct measurement of high-mass diffraction covering the mass range $3.4\ \text{GeV} < M_X < 15.7\ \text{GeV}$.
However, at this energy, ATLAS measured the distribution of rapidity gap size in minimum bias events \cite{ATLAS:2012djz}. 
For the largest measured gap sizes, \textit{i.e.} the lowest masses, the data show an approximately constant distribution of about 1 mb/unit of rapidity.

Following the argument of \cite{lowmassnote}, we now aim at estimating the rapidity span of the high-mass diffraction to be subtracted.
TOTEM T2 detectors measure up to $\eta = 6.5$, which defines the lower edge.
The upper edge is less straightforward.
The fiducial inelastic cross sections mentioned above were measured using detectors with fixed $\eta$ range, but were unfolded to correspond to a fixed limit in $M_X$.
Using the relation between the rapidity gap size and the diffractive mass, $\Delta y = -\ln M_X^2/s$, the mass limit of $M_X = 15.7$ GeV corresponds to the upper edge of $\eta = 3.28$.
However, as discussed earlier in the context of the TOTEM measurement, the relation between $\eta$ and $M_X$ is not sharp but smeared.
Fig. 1 of \cite{CMS:2012gek} shows that the acceptance of the measurement reaches $\xi$ values below $10^{-6}$, which corresponds to $M_X$  below 7~GeV ($\xi=M_X^2/s$).
Acceptance of 50\% is seen around $\xi = 1.25\cdot10^{-6}$ ($M_X = 7.8$ GeV), which can be translated into $\eta = 4.67$.
We thus attribute the uncertainty on the upper edge to $4.67 - 3.28 = 1.39$, resulting with the rapidity span of $3.22 \pm 1.39$.

Extrapolating the ATLAS rapidity gap distribution and integrating it over the above rapidity range gives the prediction of the high-mass diffraction.
Using a flat extrapolation, this results in $3.41 \pm 1.47$ mb.
Observing that the rapidity gap distribution shows a small rise for the largest gaps, one can also use a linear function to extrapolate, which gives $4.54 \pm 1.96$ mb.
The final value is thus $\sHM = 4.22\pm2.28$~mb and it leads to $\sLM = 6.8 \pm 3.1$~mb.

It should be mentioned that CMS also measured the distribution of rapidity gap size \cite{CMS:2015inp}.
Here, the gap is defined differently: for ATLAS, it starts from $|\eta|=4.9$, while for CMS it is 4.7.
In addition, the CMS measurement reaches larger gap sizes, which means shorter extrapolation to the region of interest would be needed if using the CMS data.
Unfortunately, we were not able to find the numerical results of the measurement.
On the other hand, Fig. 15 of \cite{CMS:2015inp} indicates that the distributions from ATLAS and CMS have basically the same shape, but CMS is about 20\% above ATLAS.
One can use the above information to appropriately scale and shift the fits to the ATLAS data and estimate the extrapolation of the CMS data.
The obtained value of $\sHM$ is within the estimated uncertainty.

\begin{figure}[htpb]
    \centering
    \includegraphics[width=\plotwidth]{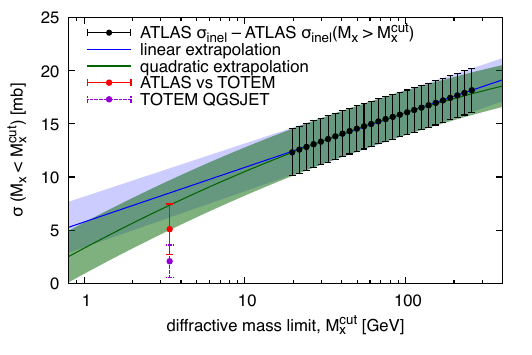}
    \caption{Diffractive cross section at $\sqrt{s} = 7$~TeV as a function of $M_X$ cut.
    The errors on the TOTEM QGSJET values are drawn with a dashed line to indicate that they are not intrinsic uncertainty of the model, but an arbitrary assumption.
    }
    \label{fig:gap}
\end{figure}

In the above-mentioned paper \cite{ATLAS:2012djz}, ATLAS translated the gap-size distribution into inelastic cross section excluding diffractive processes with a given mass cut.
These data can be directly extrapolated to the TOTEM T2 mass limit, providing an alternative to the reasoning based on \cite{lowmassnote}.
The results are presented in Fig.~\ref{fig:gap}.
The linear extrapolation to $M_X=3.4$~GeV results in $\sLM$ of 8.50 ± 2.31 mb.
A linear function corresponds to the $1/M_X^2$ behaviour of the diffractive differential cross section, which is expected to break down at low masses.
For this reason, we also apply a quadratic extrapolation, which gives 7.34 ± 2.34 mb.
Since the dominant uncertainty for the data originates from luminosity, both extrapolations were performed assuming a full correlation between the uncertainties on individual points.
For the final value, we take 7.9 ± 2.9 mb, which covers the uncertainty range of both approaches.

\begin{figure}[htpb]
    \centering
    \includegraphics[width=1.25\plotwidth]{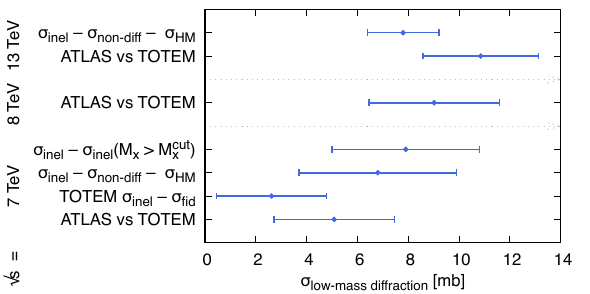}
    \caption{Summary of data-driven estimates on low-mass diffraction at LHC energies.}
    \label{fig:datasummary}
\end{figure}

One can compare the extrapolation of the ATLAS data with the low-mass cross section value used by TOTEM and with the value obtained from the ATLAS--TOTEM difference.
However, some caution is needed here, because the TOTEM T2 measurement corresponds to a sharp $\eta$ range, and not a sharp $M_X$ range.
The corresponding value of $M_X = 3.4$~GeV was obtained as the point of acceptance being close to 50\%.
This does not guarantee that $\sLM(|\eta| > 6.5) = \sLM(M_X < 3.4)$.
On the other hand, for any realistic value of the corresponding uncertainty, the ATLAS extrapolation would agree with our estimation.

All the data-driven estimates of \sLM derived in this and the previous section are presented in Fig. \ref{fig:datasummary}.
One can see that the estimate based on the difference between ATLAS and TOTEM is in good agreement with all other estimates.

\section{Conclusions}
The purpose of this article was to investigate the tension between the ATLAS and TOTEM measurements of total cross section.
The starting point was to trust the experimental groups and assume that all measurement procedures of both ATLAS and TOTEM are correct.
This led us to question the assumption about the low-mass diffraction used by TOTEM in the normalisation procedures in their main method.
 
 We showed that assuming no experimental problems with the ATLAS and TOTEM data, makes it possible to determine the cross section for the low-mass diffractive dissociation from the observed discrepancy.
 Four different approaches were considered, each based on a different set of input observables.
The resulting values were almost identical, both in the value as well as uncertainty, illustrating the robustness of the procedure.

\begin{figure}[htpb]
    \centering
    \includegraphics[width=\plotwidth]{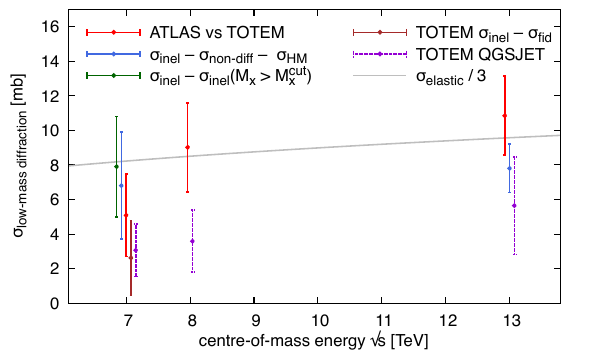}
    \caption{Summary of results on low-mass diffraction.
    The errors on the TOTEM QGSJET values are drawn with a dashed line to indicate that they are not intrinsic uncertainty of the model, but an arbitrary assumption.
    }
    \label{fig:summary}
\end{figure}

In order to validate our assumption and results, we exploited other LHC measurements sensitive to low-mass diffractive dissociation.
We offered one independent estimation of the low-mass diffractive cross section at 13 TeV, and two ways at 7~TeV.
The values based on the ATLAS--TOTEM discrepancy are in good agreement with all those points, as well as the TOTEM dedicated measurements at 7 TeV.
On the other hand, the TOTEM QGSJET value has a tendency to be below other points.
Figure \ref{fig:summary} presents a summary of all the information considered in the paper.

The fact that TOTEM QGSJET underestimates (or is at the edge of the error bars of) almost all data-driven values, with the only exception being the TOTEM measurement at 7 TeV, strongly supports the correctness of our hypothesis.
Assuming low-mass diffraction cross section above TOTEM QGSJET, including the arbitrary 50\% uncertainty, nicely agrees with all data we are aware of and solves the discrepancy between ATLAS and TOTEM total cross sections.
Given the large error bars on all the points, one cannot firmly exclude the possibility that the source of the discrepancy is partially also on the ATLAS side.
However, the results presented make it more likely that the problem is on the TOTEM side.

\section*{Acknowledgments}

We would like to express our gratitude to J. Chwastowski, V. Khoze and M. Ryskin for their useful comments on the manuscript.
This work is supported in part by the National Science Center (NCN), Poland, SONATA BIS grant no. 2021/42/E/ST2/00350. 

\appendix
\FloatBarrier
\section{Derivation of correction formulae}
\label{sec:deriv}
\renewcommand{\M}{^\ast}
Let \stot be the true value of the total cross section.
This value would be the result of the measurement if a proper correction $\epsilon$ was taken.
Taking the correction value of $\epsilon_0$ gives a value of $\stot\M$.
Applying Eq. (\ref{eq:lumiind}) to both cases and taking the ratio results in:
\begin{equation}
\label{eq:master}
\frac{\stot\M}{\stot} = 
\frac{\Nel + \Nvis (1+\epsilon)}{\Nel + \Nvis (1+\epsilon_0)}
\end{equation}
where \Nvis is the number of inelastic events within the acceptance of the detector.
Temporarily introducing integrated luminosity $L$ such that $\Nel = L (\stot - \sinel)$ and $\Nvis (1+\epsilon) = L \sinel$ results in:
\begin{equation}
\frac{\stot\M}{\stot} = 
\frac{\stot - \sinel + \sinel}{\stot -\sinel + \sinel (1+\epsilon_0) / (1+\epsilon)}.
\end{equation}
Neglecting non-linear $\epsilon$ and $\epsilon_0$ terms allows writing the above as:
\begin{equation}
\label{eq:A}
\epsilon = \epsilon_0 + \frac{\stot\M - \stot}{\sinel},
\end{equation}
which is our formula A.

Alternatively, one can observe that
\begin{equation}
\frac{\Nvis(1+\epsilon_0)}{\Nel} = \frac{\sinel\M}{\stot\M - \sinel\M}
\end{equation}
Extracting \Nvis and substituting it in Eq. (\ref{eq:master}) gives:
\begin{equation}
\label{eq:B}
\epsilon = \epsilon_0 + \frac{\stot\M(\stot\M - \stot)}{\stot \sinel\M},
\end{equation}
which is our formula B.

Formula C is obtained from formula A by substituting \stot with $\sinel / (1 - R)$, where $R$ is the ratio of the elastic to total cross section.

Formula D is derived with an additional physics assumption or neglecting non-exponential terms in the differential elastic cross section.
Then, elastic cross section can be related to \stot and the elastic slope.
This allows substituting \sinel in Eq. (\ref{eq:A}) with $\stot - \stot^2 (1+\rho^2) / (16\pi B)$, resulting in formula D.

\section{Error propagation}
\label{sec:errorprop}

Approaches A--D for calculating $\epsilon'$ use in total 6 input values:
$\stot^\md$, $\stot^\mi$, $\sinel^\mi$, $\sinel^\md$, $R^\mi$, and $B^\mi$,
each of them having some experimental uncertainty.
The uncertainties of all inputs are propagated through the appropriate $\epsilon'$ formula and added quadratically.
This procedure neglects correlations which are present between different inputs.

For example, one of the main uncertainties in the ATLAS measurements originates from the luminosity uncertainty.
It affects both $\stot^\mi$ as well as $\sinel^\mi$.
Since formula A contains the ratio of these two values, the luminosity uncertainty should, at least partially, cancel.
Neglecting the correlations leads to some overestimation of the uncertainty.
On the other hand, the luminosity uncertainty does not affect the $B$ value, and it partially cancels also in $R$.
Thus, approach C and D will be less sensitive to this correlation.
This is also true for approach B, where only a single input from the ATLAS measurement is used.
This illustrates that considering several approaches allows evaluating the effect of neglecting correlations between uncertainties.

The uncertainties on $\epsilon$ originate only from $\epsilon'$, \textit{i.e.} $\epsilon_0$ should be treated as a value that was used to obtain $\stot^\md$ and any uncertainty assigned to $\epsilon_0$ is not relevant here.
What is more, the uncertainty of $\epsilon_0$ that TOTEM propagated to their total cross section value is also not applicable for the present considerations, and including it would result in overestimation of the uncertainties on $\epsilon$.
Therefore, the $\stot^\md$ uncertainties used in the calculations shound have this contribution subtracted.

In order to understand how the uncertainty assigned to $\epsilon$ propagates to $\stot$, one can consider how \stot changes with $\epsilon$:
\begin{equation}
    \stot
    \quad
    \xrightarrow{\ \epsilon_0\ \to\ \epsilon_0 \pm \epsilon_0/2\ } 
    \quad
    \stot \mp \epsilon_0\sinel/2.
\end{equation}
Then, this contribution is quadratically subtracted from the TOTEM's model-dependent \stot uncertainties.
The results are presented in Table \ref{tab:subtraction}.

\begin{table}[htbp]
\caption{Ingredients in subtracting the extrapolation uncertainty from the TOTEM $\stot$ uncertainty.
$\delta^\md_\text{tot}$ is the uncertainty on $\stot^\md$, 
$\delta_\text{tot}^{\md,\epsilon_0}$ is contribution to $\delta^\md_\text{tot}$ related to uncertainty assumed on $\epsilon_0$,
$\delta_\text{tot}^{\md,s}$ is the $\delta^\md_\text{tot}$ with $\delta_\text{tot}^{\md,\epsilon_0}$ subtracted.
}
\label{tab:subtraction}
\centering
\vspace{3pt}
\begin{tabular}{cccccc}
\toprule
       $\sqrt{s}$ &   $\stot^\md$ [mb] &     $\delta^\md_\text{tot}$ [mb] & $\epsilon_0$ &   $\delta_\text{tot}^{\md,\epsilon_0}$ [mb] &   $\delta_\text{tot}^{\md,s}$ [mb]\\
\midrule                                                         
 13 TeV &              110.6 &                 3.4 &             0.071 &                      2.8 &                    1.9 \\
 8 TeV  &              101.7 &                 2.9 &             0.048 &                      1.8 &                    2.3 \\
 7 TeV  &               98.0 &                 2.5 &             0.042 &                      1.5 &                    2.0 \\
\bottomrule
\end{tabular}
\end{table}

\printbibliography

\end{document}